\documentclass[aps,prl,twocolumn,groupedaddress,showpacs,amsmath,amssymb]{revtex4-1}
\usepackage{epsfig}
\usepackage{amssymb}
\usepackage{amsmath}
\usepackage{graphicx}
\usepackage{amsfonts}
\usepackage{epsfig}
\usepackage{revsymb4-1}
\usepackage{bm}
\usepackage{latexsym}

\begin{document}

\title{Anomaly of the Internal Friction in the Helium Crystals Grown in the Burstlike Growth Mode}

\author{V. L. Tsymbalenko}
\affiliation{NTK Superconductivity, Russian Research Centre Kurchatov Institute, 123182 Moscow,
Russia}

\affiliation{Kapitza Institute for Physical Problems, Russian Academy of Sciences, 119334 Moscow,
Russia}

-
\begin{abstract}
The internal friction in the crystals grown in two modes, namely, slow and anomalously fast (burst-like
 growth) modes, is measured in the temperature range 0.49-0.75~K at a frequency of about 75~kHz.
An additional contribution to the damping decrement and softening of the dynamic modulus are detected,
and their relaxation to equilibrium values at a time constant of about 3~ms is observed. Possible origins of this
effect are discussed.
\end{abstract}


\maketitle

\section{INTRODUCTION}
\par
Below the second roughening temperature $(T~<~T_{R2}\approx0.9 K)$, the helium crystals having appeared in a metastable superfluid grow in two qualitatively different modes depending on initial supersaturation $Dp_0$ \cite{T1,T2}.
Until the deviation from equilibrium is rather low, the crystal face growth rate is low and determined by the
well-known mechanisms, namely, growth on screw dislocations and Frank-Read sources \cite{T3}. However,
the growth rate increases jumpwise above a certain critical supersaturation $p_c$, which depends on
temperature \cite{T4}, and a crystal grows completely in $200-400 \mu s$ \cite{T5}.
This effect is characterized by a time lag between the nucleation of a crystal and a jumplike increase in
its growth rate \cite{T6}. The crystal grows at an ordinary rate in a rather long time, which also depends on the
initial supersaturation and reaches $0.2 s$, and its growth then accelerates. At low temperatures, the growth rate
increases by a factor of $100-1000$. To date, the conditions under which the second growth scenario (burst-like
 growth) occurs have been determined \cite{T4}, the burstlike growth rates have been measured \cite{T2,T6}, and
the return to the normal state with the low growth rates typical of equilibrium crystals has been studied \cite{T7}.
\par
However, the entire set of experimental data did
not provide an unambiguous answer to the question of
whether this phenomenon is the transition of a surface
into a new rapidly growing state or is caused by a
change in the volume properties of a crystal \cite{T2}. Since
the face growth kinetics changes, it is reasonable to
assume that the effect is caused by a surface phase
transition or the appearance of a qualitatively new
intense growth source. As follows from filming the
crystal growth at the stage of burstlike growth \cite{T5}, the
anisotropy of kinetic faceting is low, which means
simultaneous acceleration of the growth kinetics of all
faces (including various orientations). This also means
that the preparation for the beginning of burstlike
growth should take the same time. Thus, the new
growth mechanism should have a low sensitivity to the
surface parameters such as the roughening temperature, the surface energy, and the interplanar spacings.

The difficulties in searching for such a mechanism
led to assumption that the growth kinetic is accelerated due to a change in the internal state of the crystal.
Only one experiment related to the volume parameters
has been performed to date: researchers tried to detect
a thermal effect during the formation of a rapidly
growing state. At temperatures above 0.4 K, no specific features were revealed during a transition into the
burstlike growth phase against the background of temperature oscillations accompanying crystal growth.
These results only give the upper estimate of the volume transition energy \cite{T8}.
\par
A change in the internal state of the crystal can
affect its kinetic properties, such as thermal conductivity and internal friction. The intrinsic dissipation
losses of an equilibrium crystal during vibrations at a
frequency of 15-80 kHz are very high \cite{T9,T10}. The
maximum damping decrement of helium is 0.5. The
temperature dependences of the dynamic modulus
and the decrement and the effect of $^3$He impurities,
plastic deformation, and annealing on internal friction
suggest that the dissipation mechanism has a dislocational nature \cite{T10}.
Since the internal friction is determined by the sample volume, we can expect that the
volume changes resulting in burstlike growth will
affect the internal friction parameters of the crystal.
Then, the return to the normal state with slow growth
kinetics should be accompanied by the attenuation of
the addition to the decrement in time.
\par
The purpose of this work is to measure the internal
friction at a frequency of about 75 kHz after crystal
growth in the normal and anomalously fast (burstlike)
growth modes. Preliminary results are published in \cite{T11}.
\par

\section{EXPERIMENTAL}
\par
At the initial time, the liquid in a container was
under the pressure exceeding the phase equilibrium
pressure by $Dp_0$. The value of $Dp_0$ was limited by spontaneous nucleation of a crystal on the wall, so that the maximum supersaturation was 5-12~mbar in various
experiments. Before the onset of growth, the pumping
of a quartz resonator was switched off and the amplitude of its vibrations began to decrease. Then, a high
voltage pulse was applied to a needle located so that
the formed crystal touched a quartz oscillator. The
additional electrostatic pressure stimulated the formation of a nucleation center for a solid phase \cite{T6}, and
the metastable liquid began to solidify. The crystals
grown by this technique have a dislocation concentration of $10^5-10^6~cm^{-2}$ [3]. Crystal growth was filmed
with a video camera. The switching-off signal from the
generator was used to start the recording of quartz
vibration amplitude and the readings of a capacitance
pressure sensor at a step of $1~\mu s$. The total recording
time was 130~ms. The frequency and $Q$ factor of the
composite system were determined from 1-ms recording segments of vibration amplitude relaxation. The
details of the optical technique were described in \cite{T6,T12}.
\begin{figure}
\includegraphics{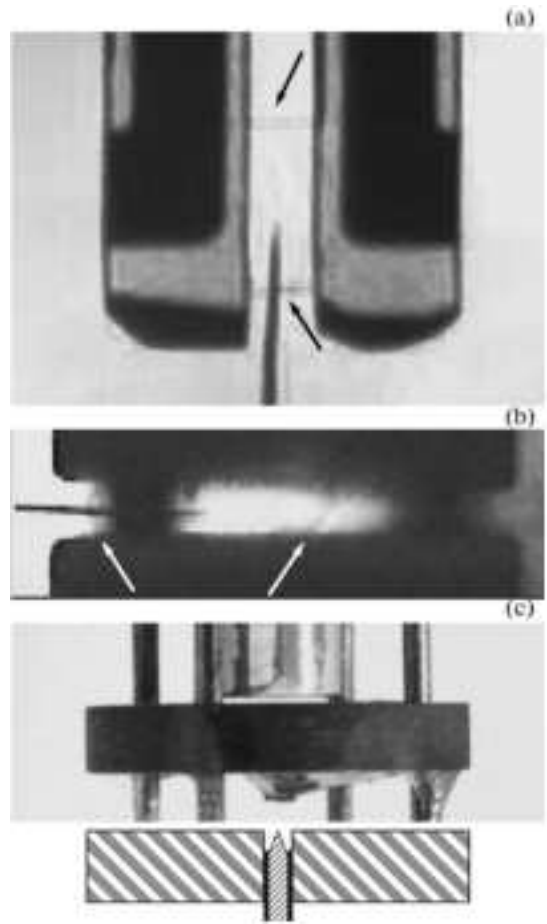}
\caption{Types of quartz resonators and the position of the
tungsten tip initiating crystal growth. The arrows in (a) and
(b) indicate crystal boundaries. (a) Crystal in the gap of a
quartz fork, (b) crystal growth in a gap of 0.4~mm, and (c)
assembly with central location of a tip and a gap of
0.18~mm (photograph and schematic section).}
\label{Fig1}
\end{figure}
\par
Figure 1 shows the quartz resonators used in experiments and the position of needles. The quartz fork
had a resonance frequency of 31909.2~Hz and an
intrinsic damping decrement $\delta_0\approx 10^{-4}$ in superfluid
helium (Fig.1a). Cylindrical quartz 3.4 mm in diameter and 28 mm in length with a fundamental torsional
vibration mode in helium was excited at a frequency of
74 551.9~Hz (Figs.1b,1c). The Q factor of free vibrations was $3.2\times10^4$. The quartz vibration amplitude was
chosen so that the maximum shear strain of a helium
crystal did not exceed $10^{-7}$. This resonator was used
earlier to study the internal friction in crystalline
helium \cite{T9,T10}.
\par
The helium single crystal having grown in the gap
during torsional vibrations undergoes complex deformation, so that its torsional rigidity is determined by
all elasticity tensor components. To process the experimental data, we use a simplified model: a helium crystal is considered as an isotropic medium. The crystal is
assumed to have the shape of a disk with its axis coinciding with the axis of torsional vibrations. In this
approximation, the equation describing the vibrations
of the quartz-helium crystal system has the form
\begin{equation} \label{e01}
\frac{\tan(k_0 L_0)}{k_0 L_0}=\left (\frac{S}{S_0}\right)^2   \frac{\cot(kh)}{kh},
\end{equation}
where $L_0$ is the quartz crystal length and $S_0$ and $S$ are
the quartz end face area and the helium crystal area,
respectively. Helium crystal area $S$ in contact with the
end face after the end of growth was estimated from
crystal volume $V_c$ by the relationship
\begin{gather}\label{e02}
V_c=V_0\frac{\rho}{\Delta\rho} k_l Dp_0=Sh,
\end{gather}
where $V_0$ is the inner container volume, $\rho$ is the liquid
helium density, $\Delta\rho$ is the difference between the solid
and liquid helium densities, $k_l$ is the compressibility
factor of liquid helium, and $h$ is the gap size. Wavevectors $k_0$ and $k$ of the torsional vibrations of quartz and
helium are described by the conventional relations
\begin{gather}\label{e03}
k_0=\frac{\omega}{c_0},     k=\frac{\omega}{c},    c^2=\frac{G'+iG''}{\rho'}
\end{gather}
where $c_0$ and $c$ are the torsional vibration velocities of
quartz and the helium crystal, respectively; $\omega$ is the
system vibration frequency; $G = G' + iG''$ is the complex dynamic shear modulus of helium; and $\rho'$ is the
solid helium density. The imaginary part of the frequency $\omega''=\omega'\delta/2\pi$ was determined from the experimental values of the frequency and damping decrement ä of the system, and the complex frequency was
then substituted into Eq.(\ref{e01}). By solving the equation,
we found the real and imaginary parts of the dynamic
modulus of helium. The decrement of the medium was
determined using the well-known relation $\delta_{He} =\pi G''/G'$. Using the simplifications described above, this
processing give results averaged over all elastic constants of the helium crystal.

In earlier internal friction experiments \cite{T9,T10}, we
used a relation deduced from energy considerations to
find $\delta_{He}$. Under the conditions $\delta \ll 2\pi$ and $\delta_0 \ll \delta$, the
decrement of the medium is calculated by the formula

\begin{equation}\label{e04}
\delta_{He} \approx  \frac{\delta - \delta_0}{1-(\omega_0/\omega)^2} \alpha,
\end{equation}
where $\omega_0$ is the vibration frequency of free-standing
quartz and $\alpha$ is the coefficient that takes into account
the inertia of the helium crystal. This coefficient is $\alpha\approx0.9$ for the given geometry and a gap of 0.4~mm. The decrements calculated by Eq.(\ref{e04}) and Eqs.(\ref{e01}) and (\ref{e02}) agree with each other to within 10\%.
\par
The growth and the size oscillations of rapidly
growing crystals end in 0.5-3~ms depending on temperature \cite{T5}. Then, the crystal volume increases due to
the flow from an external system, and this increase is
only 2\% in 20~ms. Therefore, the contact area $S$ is
assumed to be constant in approximately 2~ms and can
be calculated by Eq.(\ref{e02}). When calculating the modulus and the decrement of slowly growing crystals, we
took into account the change in their volume determined from the time dependence of the pressure decrement
in the container. Contact area $S(t)$ increased in time. Note that the estimation of $S(t)$ by Eq.(\ref{e02}) gives
a high error at the beginning of growth, when the crystal size is small or comparable with the gap size.

\section{EXPERIMENTAL RESULTS}
\par
\subsection{Experiments with the Quartz Fork}
\par
In the experiments with the fork, spontaneous
nucleation of the crystal on the container walls limited
the starting supersaturation, which was at most 5~mbar.
During crystal growth, the system frequency increased
from 32 to 60~kHz. As follows from Eq.(\ref{e03}), the $Q$ factor of the system was mainly determined by the crystal
with a decrement of 0.1-0.5. As a result, the vibrations
of the fork-crystal system decay in 1-2~ms. In this
time, the anomalous state has no time to form at such
a low supersaturation \cite{T6}. These experiments did not
lead to the desired results.
\par
\subsection{Experiments with Cylindrical Quartz}
\par
In contrast to the experiments with the fork, the
shift of the quartz-helium system frequency and the
decay of quartz are low. Free vibrations of the system
last the time that is sufficient for the anomalous state
to form and the evolution of its internal friction to be
studied during 130~ms. In the first series of experiments, a needle was placed from one side and its tip
was near the axis of quartz vibrations (Fig.1b). Experiments were carried out at temperatures of 0.49, 0.54,
0.59, 0.62, 0.65, 0.69, and 0.75~K to an initial supersaturation of 10~mbar. The scatter of the temperatures
inside each group of measurements was 0.002~K. The
gap between the end face of quartz and the metallic
base was 0.4~mm. In the second series of experiments,
a needle was located from below along the axis of torsional vibrations (Fig.1c). Experiments were performed at a temperature of 0.512~K, and the gap in this
case was 0.18~mm. The position of the needle ensured
the location of a crystal that corresponds better to the
simplified model of calculating the crystal parameters.
\par
The averaged shear modulus calculated from the
volume compressibility and the Poisson ratio ($\nu= 0.3$)
is $1.22\times 10^8$~dyn/cm$^2$. The experimental values fall in
the range $(1-3.5)\times10^8$~dyn/cm$^2$. This difference in
the results is related to the crystal anisotropy and the
simplified approach to determining the shear modulus. In Eq.(\ref{e01}), we did not take into account the presence of a rigid needle and the shift of the crystal from
the axis of quartz vibrations. The last factor can be
taken into account by introducing a coefficient, which
depends on the ratio of the crystal radius to the shift,
into Eq.(\ref{e01}). This coefficient is unity for a zero shift.
When the shift of the crystal is not taken into account,
its torsional rigidity is underestimated and, hence, the
shear modulus calculated by Eq.(\ref{e01}) is overestimated.
The same effect is caused by a needle located from one
side. This note belongs to the elastic modulus rather
than to its time dependence, since the crystal size and
shape change insignificantly in the measurement time.
The decrement of a helium crystal calculated by
Eq.(\ref{e01}) or Eq.(\ref{e04}) has a weak sensitivity to the shift. A
numerical simulation showed that the shift of the crystal from the vibration axis even by its radius changes
the decrement calculated by Eq.(\ref{e04}) by at most 10\%.

\subsection{Slow Crystal Growth}
\par
Figure 2 shows the changes in the pressure in the
container and the frequency and decrement of the
quartz-crystal system at T = 0.75~K and an initial
supersaturation of 2.7~mbar. During crystal growth,
the pressure in the container decreases in approximately 50~ms. The crystal radius calculated in approximation (\ref{e02}) increases and then, after a time of $\sim20$~ms,
changes weakly. The frequency and decrement of the
quartz-crystal system increase monotonically. The
results of calculating the decrement and the shear
modulus of a helium crystal are also depicted in Fig.2.
\begin{figure}
\includegraphics{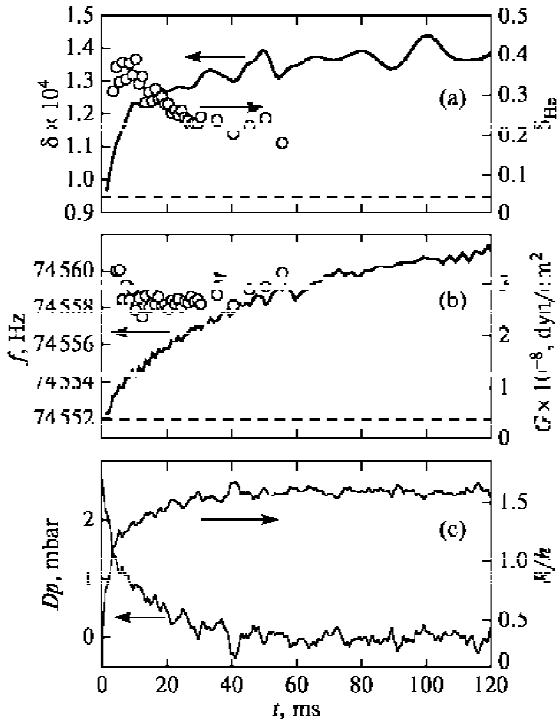}
\caption{Time evolution of the quartz-helium crystal system
parameters and the calculated values of (a) damping decrement and (b) shear modulus at T = 0.75~K in the normal
growth region. (c) Change in the pressure in the container
and the crystal radius calculated by Eq.(\ref{e02}). (b) Recorded
frequency of a quartz-helium crystal composite oscillator
and the shear modulus calculated by Eq.(\ref{e01}). (a) Change in
the damping decrement of the system and calculated decrement of a helium crystal.}
\label{Fig2}
\end{figure}
It is seen that the shear modulus falls in the range
$(2.5-3.5) \times 10^8$ dyn/cm$^2$ during crystal growth. Allowing for the simplifications and the scatter of the experimental results, we assume that the shear modulus
remains unchanged. The calculated decrement of a
helium crystal changes weakly during crystal growth
(Fig.2a). The absolute value of the decrement (0.2-0.3) agrees with the measurement results in \cite{T10}.
\par
Thus, the slow crystal growth is characterized by a
constant shear modulus and an unchanged, or weakly
changed, decrement, which is close to the decrement
of the crystals grown and annealed near the melting
curve \cite{T10}. The growth from a metastable liquid
weakly affected the internal friction of helium crystals.

\subsection{Burstlike Crystal Growth}
\par
In the range of burstlike growth, the pressure
detected by the pressure sensor decreases sharply after
the nucleation of a crystal and exhibits a jumplike
increase in the growth kinetics of all crystal faces. The
decrement and frequency of the quartz-helium system were determined from recording free vibrations
after completed crystal growth.
\begin{figure}
\includegraphics{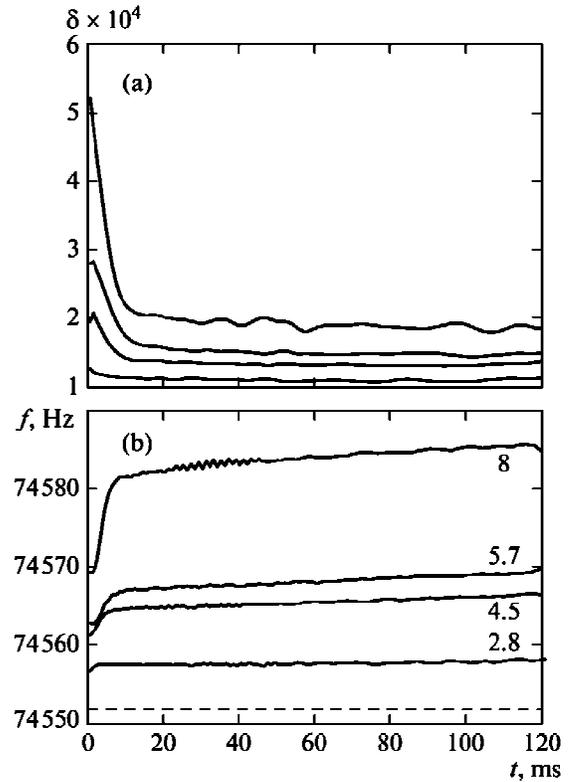}
\caption{Time evolution of (b) frequency and (a) damping
decrement of the quartz–helium crystal system during
crystal growth at various initial values of supersaturation
for T = 0.49~K after the end of burstlike growth. The
numerals in (b) indicate the values of $Dp_0$~[mbar].
The sequence of the decrement curves from bottom to top
in (a) corresponds to the sequence of the frequency curves in (b).}
\label{Fig3}
\end{figure}
Figure 3 shows the
time dependences of these parameters. The following
two regions are clearly distinguished in these curves.
After the end of crystal growth, the decrement of the
system begins to decrease simultaneously with an
increase in the averaged dynamic shear modulus. In
approximately 20~ms, relaxation is completed and a
further slow increase in the frequency and the shear
modulus is observed. The second region is related to
crystal growth due to a liquid flow along a capillary
from the external system. The slopes of the $f(t)$ dependences in this region are proportional to initial supersaturation $Dp_0$, and the calculated shear modulus and
the damping decrement of the crystal are constant
within the limits of experimental error. The values of
$\delta_{He}$ in this region change from crystal to crystal and fall
in the range 0.05-0.25. For example, for the curves
presented in Fig. 3, these values are 0.074 (2.8), 0.1
(4.5), 0.13 (5.7), and 0.11 (8); in parentheses, we give
the values of $Dp_0$~[mbar]. Such a scatter was observed
earlier in annealed samples grown at a constant pressure and a constant temperature gradient, and it is
likely to be related to different defect distributions in a
sample \cite{T10}. A nonuniform dislocation distribution in
a crystal was detected earlier during slow growth,
where the kinetics is controlled by defects (screw dislocations, Frank-Read sources \cite{T3}.), from the deviation of a crystal shape from a regular hexagonal prism.
The results obtained at the final stage of crystal growth
agree with the well-known data.
\par
An excess decrement and a decrease in the
dynamic modulus are only observed in the crystals
grown in the burstlike growth mode. Figure 4 shows
the evolution of the decrement and the shear modulus
of a helium crystal normalized by their steady values at
$t >20$~ ms.
\begin{figure}
\includegraphics{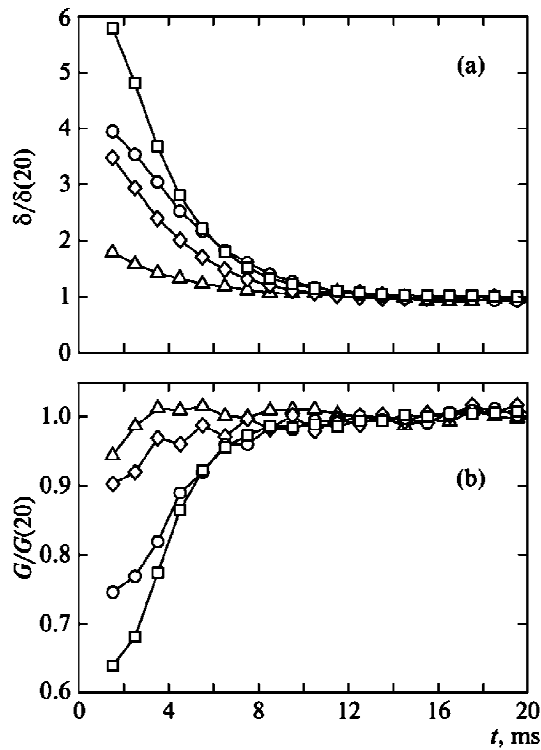}
\caption{(a) Damping decrement and (b) shear modulus calculated from the curves in Fig. 3 and normalized by their
steady values at t = 20~ms. The initial supersaturation is ($\Box$)
8, ($\bigcirc$) 5.7, ($\diamond$) 4.5, and ($\bigtriangleup$) 2.8~mbar.}
\label{Fig4}
\end{figure}
As is seen from Figs.3 and 4, the relaxation
contribution increases with the initial supersaturation.
The relaxation curves are well described by the exponential relationship
\begin{equation}\label{e05}
\frac{\delta (t)}{\delta (20)} =1 + A \exp \left( -\frac{t}{\tau}. \right)
\end{equation}
The dependences of parameters $A$ and $\tau$ on $Dp_0$ are
presented in Fig.5. The increase in the relaxation contribution is illustrated by an increase in amplitude $A$
with the initial supersaturation. Parameter $A$ vanishes
at the supersaturation that corresponds to the boundary supersaturation separating the regions of normal
and anomalous growth, $A(Dp_0 \rightarrow p_c) \rightarrow 0$ \cite{T4}.
\begin{figure}
\includegraphics{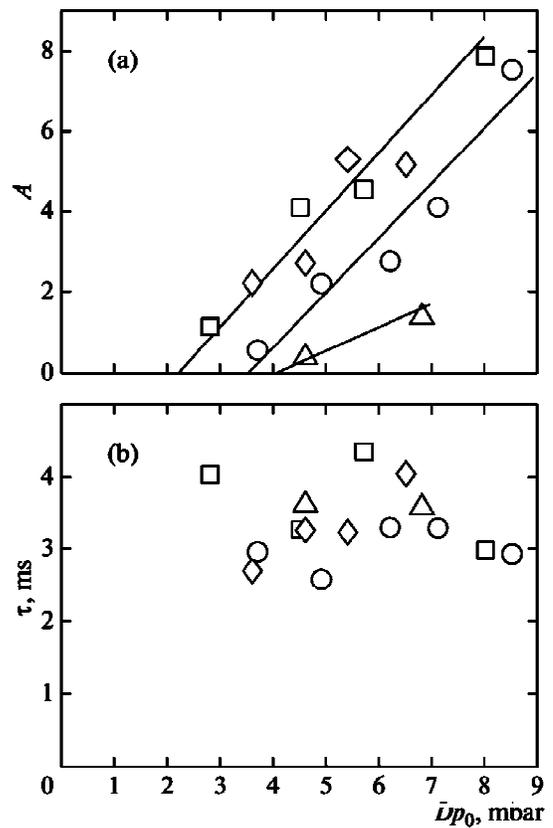}
\caption{(a) Excess decrement amplitude and (b) process
relaxation time calculated by Eq.(\ref{e05}) vs. the initial supersaturation at a temperature of ($\Box$) 0.49, ($\bigcirc$) 0.59, ($\diamond$)
0.65, and ($\bigtriangleup$) 0.69 K.}
\label{Fig5}
\end{figure}
The relaxation time falling in the range 2.5-4.5~ms depends on neither temperature nor the initial supersaturation
within the scatter of the experimental results.
\par
Thus, the crystals grown in the burstlike growth
mode exhibit an excess decrement and a decrease in
the dynamic shear modulus, which relax in time, in
the temperature and supersaturation ranges under
study. The effect is low or zero at the boundary
between anomalous and normal growth and increases
with the distance from it. This addition decreases at a
time constant of 3~ms.

\section{DISCUSSION OF RESULTS}
\par
The measured decay is caused by dissipation both
inside the crystal and at its surface. It is known from
the experimental data that high face kinetics after
crystal growth lasts at least 3~ms and that the return to
the normal slow kinetics occurs in 50-100 ms \cite{T7}. The
elastic stresses at the crystal boundary change the
chemical potential of the solid phase and initiate
interface motion. A similar effect was studied in \cite{TWE}.
As a result, melting takes place and crystal faces grow
at a double frequency, since the addition to the chemical potential is quadratic in stress. In turn, the surface-assisted dissipation consists of two parts. The first
part is caused by the surface growth kinetics, is determined by kinetic growth coefficient $K$, and is lower
than $10^{-7}$. Oscillating growth is accompanied by liquid
flows, which induce sound emission at a double frequency. The second contribution is also low, less than
$10^{-7}$. The sum of both contributions is much lower
than the decrement of helium detected in experiments, which is at least 0.07. Thus, the additional
decay is related to volume processes.
\par
The internal friction measured in experiments is
determined by the sum of all dissipation processes
occurring in the crystal. The possible sources of the
effect under study are as follows. The additional decay
can be caused by the appearance of a new relaxation
process. The time behavior of decaying agrees qualitatively with the change in the face growth kinetics after
the end of growth. Therefore, the jumplike state of the
crystal is assumed to manifest itself in both the appearance of an additional decay process and the acceleration of face growth.
\par
The second possibility consists in the effect of the
changed state of the crystal on the oscillation of dislocation segments, which are considered to be responsible for the high damping decrement in crystalline
helium. In the Granato-L\"{u}cke model of dislocational
internal friction \cite{GL}, the frequency and temperature
dependences of internal friction in the case of damped
oscillation of dislocation segments are expressed by
the formula
\begin{equation}\label{e06}
\delta = \xi \frac{4(1-\nu)}{\pi^2} \Lambda L^2 \Omega \frac{\omega \tau_{dis}}{1+(\omega \tau_{dis})^2},
 \tau_{dis} \sim B(T)L^2
\end{equation}
where $B$ is the damping constant, $L$ is the average dislocation segment length, $\Lambda$ is
the dislocation concentration, $\Omega$ is the orientation
factor, and $\xi$ is the parameter that takes into account
the length distribution of dislocation segments. For a $\delta$
distribution, we have $\xi$= 1; for a network of random
intersections, $\xi \approx 4.4$. In terms of this model, a
decrease in the decrement with a simultaneous
increase in the dynamic modulus occurs when time $\tau_{dis}$
increases, i.e., when the damping constant determined by the phonon gas viscosity increases. Whence,
it follows that the dislocation motion in the crystal
grown in the burstlike growth mode is retarded by
phonons to a lesser extent as compared to that in an
equilibrium crystal.
\par
The last conclusion holds true if the geometric configuration of a dislocation network remains unchanged
after the end of growth. However, as is known from
earlier experiments on internal friction  \cite{T10}, even slow
growth of helium crystals can lead to time variations of
the decrement and the dynamic elastic modulus of a
crystal. It is necessary to perform long-term annealing
of a helium crystal near the melting curve and to provide a slow temperature change during measurements
to obtain reproducible results. Taking this fact into
account, we assume that burstlike crystal growth creates a nonequilibrium network of dislocations and it
begins to relax after the end of growth, which is
reflected on the values of the decrement and the
dynamic modulus. For a network of dislocations with
self-intersections, we have $\Lambda L^2 \approx 1$; therefore, it
weakly changes upon annealing and the time evolution
is related to an increase in the average dislocation segment length during rapid annealing, according to
Eq.(\ref{e06}). This assumption is in conflict with the results
of burstlike growth experiments on the crystals that
have no Frank-Read sources and screw dislocations
\cite{RH1, RH2}. In these experiments, a crystal face had no
growth sources after the end of burstlike growth; that
is, the burstlike growth mode did not generate defects.
Even multiple repetition of the process did not degrade the surface.
\par
In conclusion, note that the excess decaying and
the modulus defect, which relax in time, are only
observed in the crystals grown in the burstlike growth
mode. It is impossible to find the source of the anomaly using the data obtained. The effect can be
explained by both a change in the internal state of the
crystal, which manifests itself in either the appearance
of a new relaxation channel or an action on dislocation
drag, and annealing of a nonequilibrium network of
dislocations, which appears during burstlike growth.
The question of whether burstlike growth can generate
a nonequilibrium network of dislocations remains open.
\par
\section{ACKNOWLEDGMENTS}
\par
I thank A.F. Andreev for the possibility of performing this work at the Kapitza Institute for Physical
Problems; V.V. Zav'yalov for his support; V.S. Kruglov
and V.A. Sharykin (Russian State Scientific Center
Kurchatov Institute) for their assistance; and
S.N. Burmistrov and L.B. Dubovskii for useful discussions.

\end{document}